\documentclass[prl,twocolumn,showpacs,amsmath,amssymb]{revtex4}
\usepackage{graphicx}

\begin{document}

\title{\boldmath
      Measurements of the braching fractions for 
$\psi(3770)\rightarrow D^0\bar D^0,~D^+D^-,~D \bar D$ and 
the resonance parameters of $\psi(3770)$ and $\psi(2S)$}
\author{
\begin{small}
M.~Ablikim$^{1}$,      J.~Z.~Bai$^{1}$,            Y.~Ban$^{11}$,
J.~G.~Bian$^{1}$,      X.~Cai$^{1}$,               H.~F.~Chen$^{15}$,
H.~S.~Chen$^{1}$,      H.~X.~Chen$^{1}$,           J.~C.~Chen$^{1}$,
Jin~Chen$^{1}$
% ...
\end{small}
\\(BES Collaboration)\\
}
\vspace{0.2cm}
\affiliation{ 
\begin{minipage}{145mm}
$^{1}$ Institute of High Energy Physics, Beijing 100049, People's Republic
of China\\
% ...
$^{11}$ Peking University, Beijing 100871, People's Republic of China\\
% ...
$^{15}$ University of Science and Technology of China, Hefei 230026,
People's Republic of China\\
% ...
\vspace{0.4cm}
\end{minipage}
}
 
\begin{abstract}
  We report measurements of the branching fractions for
$\psi(3770)\rightarrow D^0 \bar D^0, D^+D^-,~D\bar D$ 
and resonance parameters of  $\psi(3770)$ and $\psi(2S)$.
By analyzing the line-shapes of the cross sections for
inclusive hadron, $D^0 \bar D^0$ and  $D^+D^-$ event production
in the range from 3.660 GeV to 3.872 GeV 
covering both $\psi(2S)$ and $\psi(3770)$ resonances,
we extract the branching fractions
for $\psi(3770)$ decay into $D^0\bar D^0~{\rm and}~D^+D^-$ respectively to be
$B(\psi(3770)\rightarrow D^0 \bar D^0)=(46.7 \pm 4.7 \pm 2.3)\%$ and
$B(\psi(3770)\rightarrow D^+ D^-)=(36.9 \pm 3.7 \pm 2.8)\%$, 
which give
$B(\psi(3770)\rightarrow D \bar D)=(83.6 \pm 7.3 \pm 4.2)\%$ and
non-$D\bar D$ branching fraction of $\psi(3770)$
to be $B(\psi(3770)\rightarrow non-D \bar D)=(16.4 \pm 7.3 \pm 4.2)\%$.
We meanwhile obtain the resonance parameters of $\psi(3770)$ and
$\psi(2S)$ to be 
$M_{\psi(3770)}=3772.2 \pm 0.7 \pm 0.3$ MeV,
$\Gamma^{\rm tot}_{\psi(3770)}=26.9 \pm 2.4 \pm 0.3$ MeV
and $\Gamma^{ee}_{\psi(3770)}=251 \pm 26 \pm 11$ eV;
$M_{\psi(2S)}=3685.5 \pm 0.0 \pm 0.3$ MeV,
$\Gamma^{\rm tot}_{\psi(2S)}=331 \pm 58 \pm 2$ keV
and $\Gamma^{ee}_{\psi(2S)}=2.330 \pm 0.036 \pm 0.110$ keV; as well as
the $R$ value for the light hadron production directly through one photon
annihilation to be $R_{uds}=2.262\pm 0.054\pm 0.109$ in this
energy region.

\end{abstract}

\maketitle

     The $\psi(3770)$ resonance was discovered by MARK-I Collaboration in
analysis of $e^+e^-$ annihilation 
into hadrons 29 years ago~\cite{mark1}.
Since its mass is above open
charm-pair threshold and its width is two orders of magnitude larger
than that of the $\psi(2S)$,
it is thought to decay almost entirely to pure
$D \bar D$~\cite{Bacino}.
However, there were historically large discrepancies 
between the observed cross sections
for $D\bar D$ and $\psi(3770)$ production in the $e^+e^-$ annihilation.
The observed cross section $\sigma^{\rm obs}_{\psi(3770)}$ 
for $\psi(3770)$ production 
can be obtained based on $\psi(3770)$
resonance parameters
~\cite{pdg04} and
radiative corrections~\cite{kuraev}, which yield
$\sigma^{\rm obs}_{\psi(3770)}=8.12\pm 1.56$ nb 
at the center-of-mass (c.m.)  energy $E_{\rm cm}=3.7699$ GeV and
$\sigma^{\rm obs}_{\psi(3770)}=7.53\pm 1.44$ nb at $E_{\rm cm}=3.773$ GeV.
There are three absolute
measurements~\cite{mark3_xsctdd}\cite{cleo_xsctdd}\cite{bes_xsctdd} of 
the observed cross sections for $D\bar D$ production 
in the $e^+e^-$ annihilation nearby the peak of $\psi(3770)$, 
which are summarized in table~\ref{tbl_xsct_ddbar}.
With the observed cross section 
$\sigma^{\rm obs}_{D\bar D}=5.0\pm 0.6$ nb for $D\bar D$ production
at $E_{\rm cm}=3.768$ GeV
measured by MARK-III Collaboration, 
one historically found that about $38\%$~\cite{rong_zhang_chen}
of $\psi(3770)$ does not decay into $D\bar D$ final states.
Scaling the $\sigma^{\rm obs}_{D\bar D}$
measured at $E_{\rm cm}=3.768$ GeV 
by using PDG $\psi(3770)$ parameters~\cite{pdg04}
yields the expected-observed cross section to be 
$\sigma^{\rm scaled}_{D\bar D}=4.7\pm 0.6$ nb at $E_{\rm cm}=3.773$ GeV.
The weighted average of the three cross sections listed in the second row of
table~\ref{tbl_xsct_ddbar} is
$\bar \sigma^{\rm obs}_{D\bar D} = 6.25 \pm 0.15$ nb.
Comparing the two observed cross sections for $\psi(3770)$ and $D \bar D$
production one finds that 
$(17.0 \pm 16.0) \%$ of $\psi(3770)$ does not decay into $D\bar D$,
where the large error is dominated by the uncertainties in the resonance
parameters of $\psi(3770)$.

To understand the 
discrepancy between the 
$\sigma^{\rm obs}_{\psi(3770)}$ and
$\sigma^{\rm obs}_{D\bar D}$, one has to more precisely
measure both the $\psi(3770)$ resonance parameters and 
the $D\bar D$ cross section, search for some non-$D\bar D$ 
decays of $\psi(3770)$ and directly measure the branching fraction
for $\psi(3770)\rightarrow D\bar D$. To reduce some possible systematic
shifts in comparison of the two cross sections from different
experiments due to normalizations, 
one had better perform to simultaneously measure the
$\psi(3770)$ resonance parameters and the $D\bar D$ cross section
with a same data set.
A better way to measure the inclusive branching fraction for
$\psi(3770)\rightarrow D\bar D$
is simultaneously to analyze the energy dependent
cross sections for the inclusive hadron, $D^0 \bar D^0$ and $D^+D^-$ event
production in the energy range covering both $\psi(2S)$ and $\psi(3770)$
resonances. In this way one can also more accurately measure the resonance
parameters of both the $\psi(3770)$ and $\psi(2S)$, since they are correlated each
other in analysis of the cross section scan data.

In this Letter, we present a line-shape analysis of
the inclusive hadron, $D^0 \bar D^0$ and $D^+D^-$ event  
production, by which we extract 
the branching fractions for
$\psi(3770)\rightarrow D^0 \bar D^0, D^+D^-,~D\bar D$ 
as well as for $\psi(3770)\rightarrow {\rm non-} D\bar D$ 
for the first time,
and meanwhile measure the resonance parameters 
of $\psi(3770)$ and $\psi(2S)$ 
with improved precision on $\psi(3770)$ resonance parameters
and with precision in measuring leptonic width of $\psi(2S)$ 
comparable to the current PDG world averages.
From this analysis we also measure the $R$ value for light hadron production
directly through one photon annihilation in the energy region from 3.660 GeV
to 3.872 GeV.
The analysis is based on the data taken 
with the BES-II detector~\cite{bes2} at the BEPC Collider 
during the time period from March to April, 2003.
\begin{table}
\centering   
\caption{The observed $D\bar D$ cross sections $\sigma^{\rm obs}_{D\bar D}$
which were directly measured by
MARK-III, CLEO and BES Collaborations, where $^{\rm scaled}$ means that the
observed cross section is 
scaled from
the measured value at 3.768 GeV to that at 3.773 GeV based on PDG
$\psi(3770)$ resonance parameters.} 
\label{tbl_xsct_ddbar}
\begin{tabular}{cccc} \hline \hline
$E_{\rm cm}$ (GeV) & MARK-III~(nb)  &  CLEO~(nb)  & BES~(nb) \\ \hline 
3.768 &  $5.0  \pm 0.6$  &    &           \\
3.773 &  $4.7 \pm 0.6$$^{\rm scaled}$ & $6.39\pm 0.16$ & $5.93 \pm 0.58$ \\             
\hline \hline
\end{tabular}
\end{table}  

The observed hadronic cross section is determined by the relation
\begin{equation}
\sigma^{\rm obs}_{\rm had} =\frac{N^{\rm obs}_{\rm had}}
                {L~\epsilon_{\rm had}~\epsilon_{\rm had}^{\rm trig}
                },
\end{equation}
where $N^{\rm obs}_{\rm had}$ is the number of
the observed hadronic events,
$L$ is the integrated luminosity,
$\epsilon_{\rm had}$ is the efficiency for detection
of the inclusive hadronic events and
$\epsilon_{\rm had}^{\rm trig}$ is the trigger efficiency
for collecting the hadronic events in on-line data acquisition.

The hadronic events are required to have more than 2 good
charged tracks, each of which is required to satisfy
the following selection criteria:
        (1) the charged track must be with
            a good helix fit and the number of $dE/dx$ hits per charged
            track is required to be greater than 14;
        (2) the point of the closest approach to the beam line
            must have radius $\le 2~\rm cm$;
        (3) $|\cos \theta| < 0.84$, where $\theta$ is the polar angle
              of the charged track;
        (4)  $2.0~ns~<T_{TOF}<T_p + 2.0~ns$, where $T_{TOF}$
               is the time-of-flight of the charged particle, and
               $T_p$ is the expected time-of-flight of proton with
               a given momentum;
        (5) the charged track must not be identified as a muon;
        (6) $p<E_b+0.1\times E_b\times \sqrt {(1+E_b^2)}$, where $p$ is
            the charged track momentum and $E_b$ is the beam energy in GeV;
        (7) for the charged track, the energy deposited in the BSC should be
              less than 1.0 GeV.
In addition, the total energy deposited by an event in the BSC should be
              greater than $28\%$ of the beam energy.
Furthermore, the selected tracks must not all point 
into the same hemisphere
in the $z$ direction.

\begin{figure}
\includegraphics[width=9.0cm,height=9.5cm]
{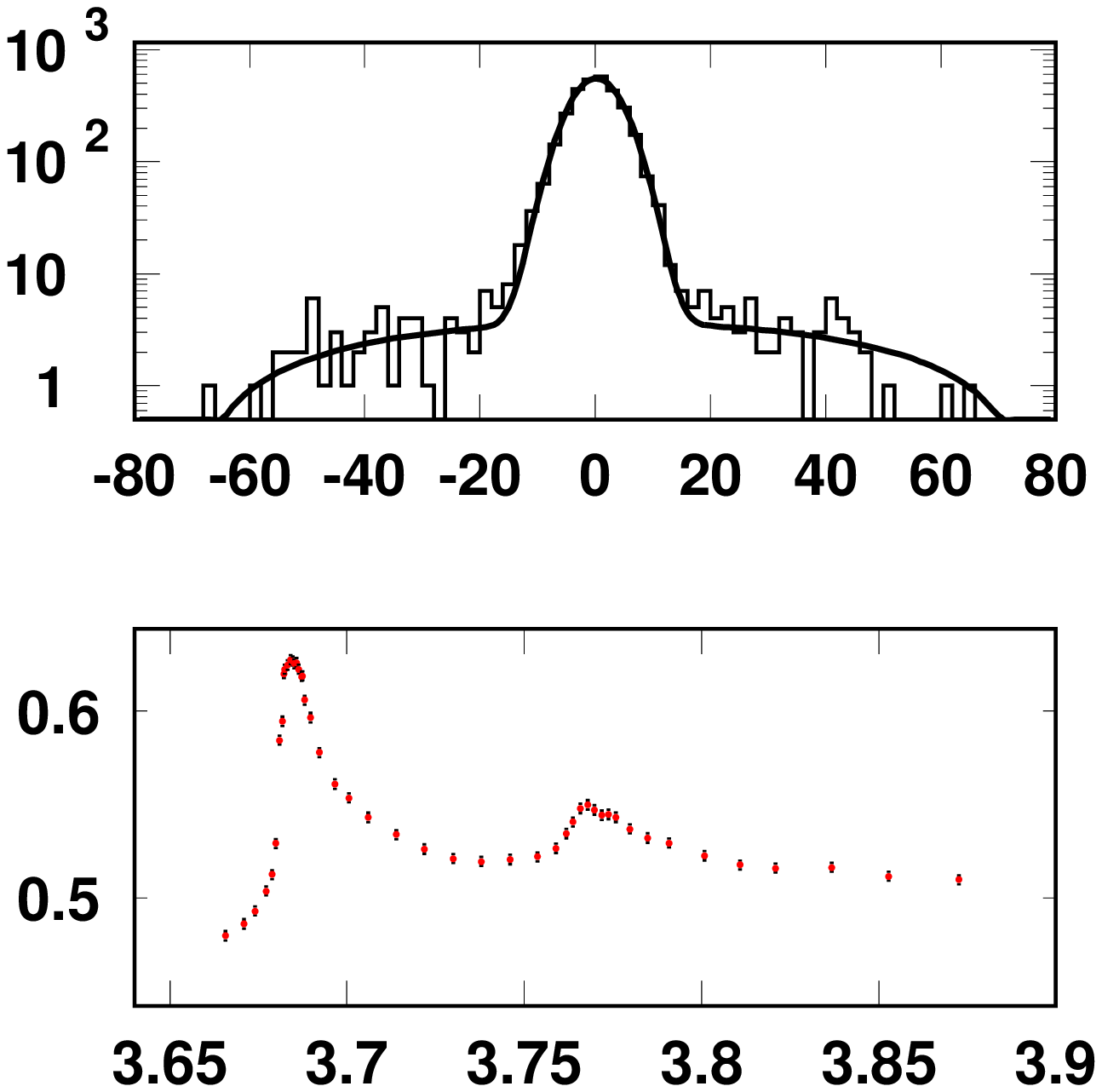}
\put(-150,130){\bf\large Averaged Z~~~~ [cm]}
\put(-150,0){\bf\large $E_{cm}$~~~~ [GeV]}
\put(-260,145){\rotatebox{90}{\bf\large Number of Events}}
\put(-260,70){\rotatebox{90}{\bf\large $\epsilon_{\rm had}$}}
\caption{(a) the distribution of the averaged $z$ of the events
satisfying the hadronic event selection criteria,
where the histogram shows the events from the data,
and the curves give the best fit to the $z$ distribution;
(b) the efficiencies for detection of the inclusive hadronic events vs
the nominal center-of-mass energies.
}
\label{Zevnt_had}
\end{figure}

  Some beam-gas associated background events can also
satisfy above selection criteria.
However, these background events
are produced at random $z$ positions in the beam pipe, while the hadronic
events are produced around $z=0$,
this characteristic can be used to distinguish the hadronic events from
the beam-gas associated background events.
Fig.~\ref{Zevnt_had}(a)
shows the distribution of the averaged $z$ of
the accepted events which satisfy the selection criteria.
Using a Gaussian function plus a second
order polynomial to fit the averaged $z$ distribution of the events,
we obtain the number, $N_{\rm had}^{\rm zfit}$, of the candidates for
hadronic events.
This number of candidates for hadronic events contains some
contaminations from
some background events such as
$e^+e^- \rightarrow \tau^+\tau^-$,
$e^+e^- \rightarrow (\gamma) e^+e^-$,
$e^+e^- \rightarrow (\gamma) \mu^+\mu^-$,
and two-photon processes. The number of
the background events, $N_{\rm b}$, due to these processes can be estimated
by means of Monte Carlo simulation. Subtracting $N_{\rm b}$ from
$N^{\rm zfit}_{\rm had}$ yields
the number of the observed hadronic events, $N^{\rm obs}_{\rm had}$.
The systematic uncertainty in
measuring the produced hadronic events
due to the hadronic event selection criteria
is estimated to be about $\sim 2.5\%$.

The integrated luminosities of the data sets are determined
using large-angle Bhabha scattering events,
which satisfy the following selection criteria:
(1) two charged tracks with total charge zero,
for each track, the point of the closest approach to the beam line must have
the  ${\rm radius} <1.5~{\rm cm}$
and $|z| <15~{\rm cm}$ where $|z|$ is measured
along the beam line from the nominal beam crossing point;
(2) each track is required to satisfy $|\cos\theta| < 0.7$,
where $\theta$ is the polar angle of the charged track,
(3) it is required that the energy deposited for
each charged track in the BSC be greater than 1.1 GeV
and at least the magnitude of one charged
track momentum be greater than  $0.9E_{\rm b}$,
(4) no track goes through the rib regions of the BSC.
The systematic uncertainty in the measured luminosities
arises mainly from the difference between the data and Monte Carlo
simulation. The total
uncertainty is estimated to be  $\sim 1.8\%$.

The detection efficiency for hadronic events is determined
via a special Monte Carlo generator~\cite{zhangdh_gen}
in which the radiative corrections to $\alpha^2$ order
are taken into account.
These generated events are simulated with
the GEANT3-based Monte Carlo simulation package~\cite{bes2_mcpck}.
The systematic uncertainty in the efficiencies due to the generator
is estimated to be $\sim 2.0\%$ ($\sim 0.7\%$) for reconstruction 
of the hadronic events from $\psi(3770)$ and 
$\psi(2S)$ decays (from continuum hadrons).
Fig.~\ref{Zevnt_had}(b) shows the Monte Carlo efficiencies
for detection of the hadronic events
produced at the different nominal center-of-mass energies.
These efficiencies are used in the measurements of the observed inclusive
hadronic cross sections at each of the energy points.

The trigger efficiencies are obtained by comparing the responses to
different
trigger requirements in the data taken at 3.097 GeV during the cross section
scan experiment. The efficiencies are measured to be
$\epsilon_{\rm trig}=(100.0^{+0.0}_{-0.5})\%$
for both the $e^+e^- \rightarrow (\gamma) e^+e^-$ and
$e^+e^-\rightarrow {hadrons}$ events.

\begin{figure}
\includegraphics[width=9.0cm,height=12.0cm]
{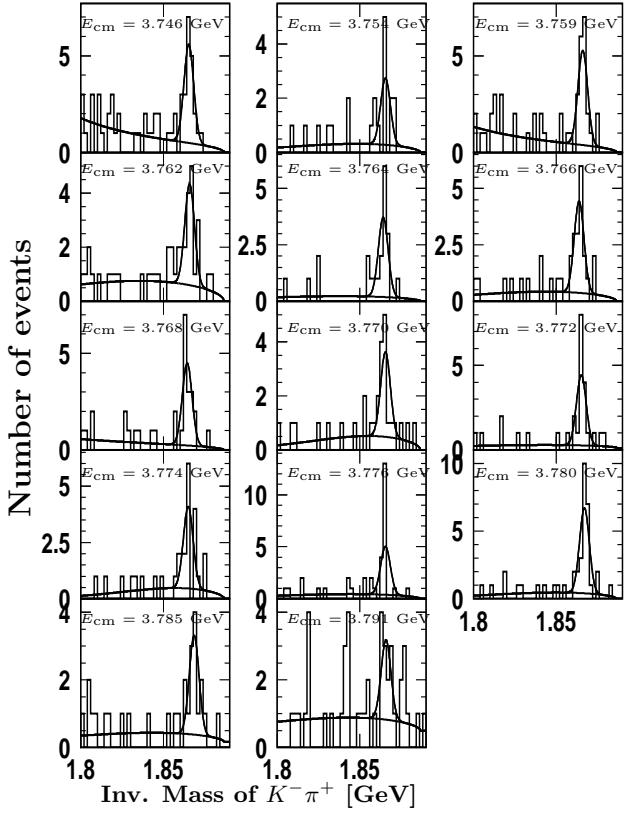}
\put(-210,22){\bf Inv. Mass of $K^-\pi^+$  [GeV]}
\put(-245,130){\rotatebox{90}{\bf\large Number of events}}
\put(-218,315.0){\tiny $E_{\rm cm}=3.746$ GeV}
\put(-140,315.0){\tiny $E_{\rm cm}=3.754$ GeV}
\put(-69,315.0){\tiny $E_{\rm cm}=3.759$ GeV}
\put(-218,260.0){\tiny $E_{\rm cm}=3.762$ GeV}
\put(-140,260.0){\tiny $E_{\rm cm}=3.764$ GeV}
\put(-69,260.0){\tiny $E_{\rm cm}=3.766$ GeV}
\put(-218,200.0){\tiny  $E_{\rm cm}=3.768$ GeV}
\put(-140,200.0){\tiny  $E_{\rm cm}=3.770$ GeV}
\put(-69,200.0){\tiny  $E_{\rm cm}=3.772$ GeV}
\put(-218,145.0){\tiny  $E_{\rm cm}=3.774$ GeV}
\put(-140,145.0){\tiny  $E_{\rm cm}=3.776$ GeV}
\put(-69, 145.0){\tiny  $E_{\rm cm}=3.780$ GeV}
\put(-218,90.0){\tiny  $E_{\rm cm}=3.785$ GeV}
\put(-140,90.0){\tiny  $E_{\rm cm}=3.791$ GeV}
\caption{The distributions of the invariant masses of the $K^-\pi^+$
combinations from the data sets taken at different nominal c.m. energy.
}
\label{sbgtg}
\end{figure}

The observed cross section for
$D^0 {\bar D}^0$ (or $D^+D^-$)
production is determined based on the number
$N_{D^0_{\rm tag}}$ (or $N_{D^+_{\rm tag}}$) of
the reconstructed $D^0$ (or $D^+$) events by

\begin{equation}
 \sigma_{D^0{\bar D}^0({\rm or~}D^+D^-)}^{\rm obs} =
                       \frac {N_{D^0_{\rm tag}} ({\rm or~}N_{D^+_{\rm tag}})
}
                       {2 \times L \times B \times \epsilon },
\end{equation}
where $L$ is the integrated luminosity of the data set used in the
analysis, $B$ is the branching fraction
for the decay mode in question,
and  $\epsilon$ is the efficiency determined from Monte Carlo simulation
for reconstruction of this decay mode.
The observed numbers, $N_{D^0_{\rm tag}}$ (or $N_{D^+_{\rm tag}}$), 
of the singly tagged $D^0$ (or $D^+$) are obtained by analyzing the invariant
mass spectra of $K^{\mp}\pi^{\pm}$ and $K^{\mp}\pi^{\pm}\pi^{\pm}\pi^{\mp}$ 
(or $K^{\mp}\pi^{\pm}\pi^{\pm}$) as
discussed in detail in Ref.~\cite{xsct_ddbar_bes}.
As an example,
Fig.~\ref{sbgtg} shows the distributions of the invariant masses of $K^-\pi^+$
combinations observed from 14 data sets collected at different energy points
in $\psi(3770)$ resonance region.

\begin{figure}
\includegraphics[width=8.0cm,height=10.0cm]
{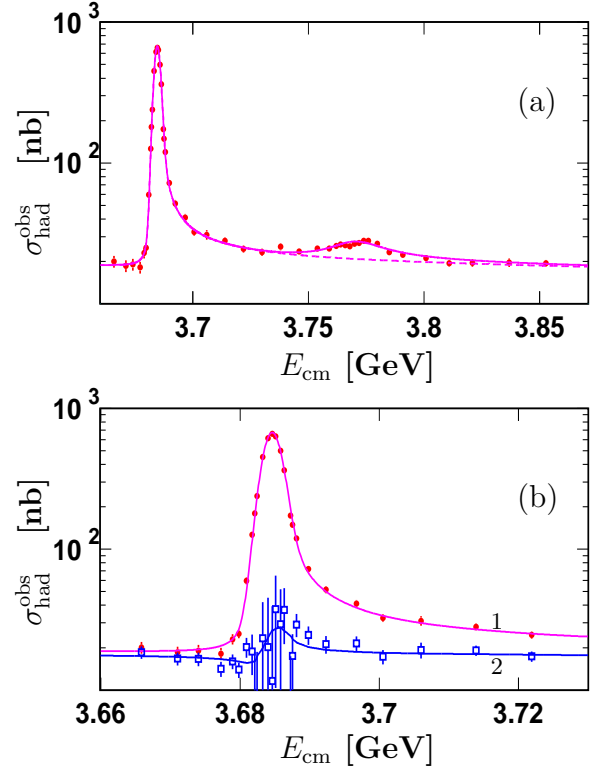}
\put(-135,135.0){\bf\large $E_{\rm cm}$ [GeV]}
\put(-135,-10.0){\bf\large $E_{\rm cm}$ [GeV]}
\put(-235,180){\rotatebox{90}{\bf\large $\sigma^{\rm obs}_{\rm had}$~~~[nb]}}
\put(-235,40){\rotatebox{90}{\bf\large $\sigma^{\rm obs}_{\rm had}$~~~[nb]}}
\put(-45,235.0){\large (a)}
\put(-45,85.0){\large (b)}
\put(-55,39.0){1}
\put(-55,22.0){2}
\caption{The hadronic cross sections versus
the nominal center-of-mass energies, where (a) the points with error show the
observed cross sections, the solid line shows the fit to the cross sections and
the dashed line represents the contributions from $J/\psi$, $\psi(2S)$ and
continuum hadron production; where (b) the points with error show the
observed cross sections and the line $1$ gives the fit to the data 
in $\psi(2S)$ resonance region, while the squares with error 
show the observed cross sections for the continuum hadron production, 
which are corrected for the radiative corrections; 
the line $2$ gives the fit to the cross sections (see text).}
\label{xsct_had_ddbar1}
\end{figure}

\begin{figure}
\includegraphics[width=9.0cm,height=9.0cm]
{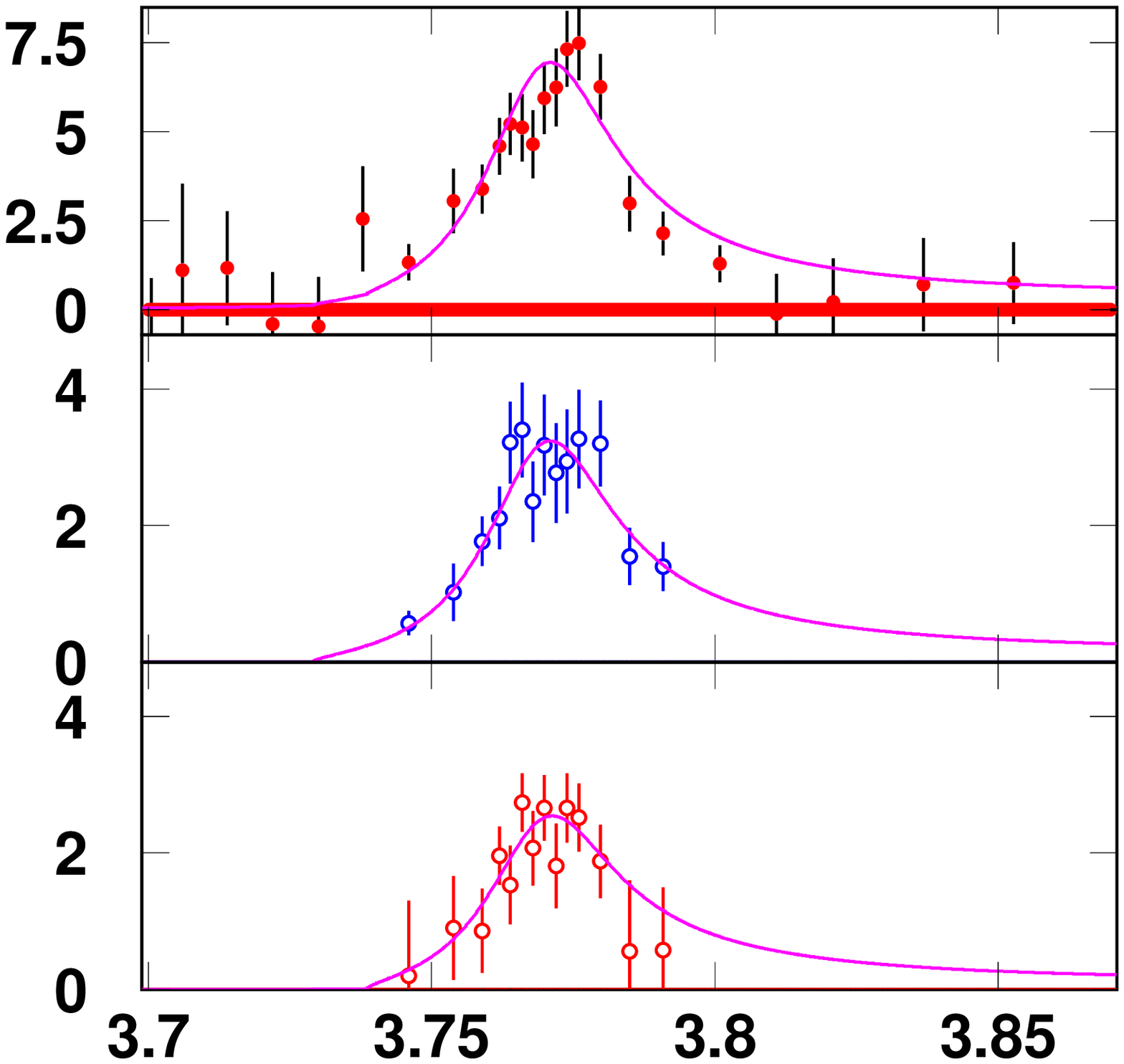}
\put(-155,-2){\bf\large $E_{\rm cm}$~~ [GeV]}
\put(-255,75){\rotatebox{90}{$\sigma^{\rm obs}_{D\bar D}$~~[nb]} }
\put(-255,155){\rotatebox{90}{\large $\sigma^{\rm obs}_{\rm had}$~~ [nb]} }
\put(-50,205.0){(a)}
\put(-50,140.0){(b)}
\put(-50,80.0){(c)}
\caption{The observed cross sections versus the nominal center-of-mass
energies, where (a) shows the inclusive hadronic event
production, (b) and (c) show the $D^0\bar D^0$ and $D^+D^-$ event
production, respectively; the points with error are the data, while
the lines are the fits to the data.
}
\label{xsct_had_ddbar2}
\end{figure}

Using the methods discussed above and in Ref.~\cite{xsct_ddbar_bes}, 
we obtain the observed
cross sections for both the inclusive hadron and $D \bar D$ event
production at the energies at which the data were collected.
Fig.~\ref{xsct_had_ddbar1} shows the observed cross sections (points with
error) for inclusive hadronic event production, while
Fig.~\ref{xsct_had_ddbar2}(b) and~\ref{xsct_had_ddbar2}(c) respectively display
the observed cross sections (points with error) 
for $D^0\bar D^0$ and $D^+D^-$ production,
where the error bars represent the combined statistical 
and point-to-point systematic uncertainties
including the statistical uncertainties in the luminosity and the
efficiencies for detection of the hadronic events and Bhabha events.

The determination of the branching fractions for
$\psi(3770) \rightarrow {D^0 \bar D^0},~D^+D^-,~{\rm and~}D\bar D$
is accomplished by simultaneously fitting the observed cross sections
for $\psi(2S)$, $\psi(3770)$, $D^0\bar D^0$ and $D^+D^-$
to functions 
that describe the combined $\psi(2S)$, $\psi(3770)$ resonance shapes,
the tail of $J/\psi$ resonance and the 
non-resonant hadronic background,
as well as the partial $\psi(3770)$ resonance shapes 
for $\psi(3770) \rightarrow {D^0 \bar D^0}$ and
$\psi(3770) \rightarrow {D^+D^-}$.
The functions are corrected for the radiative
corrections~\cite{kuraev}\cite{berends}.

For $J/\psi$ and $\psi(2S)$ resonances, we
take the Born order Breit-Wigner function
\begin{equation}
\sigma^{\rm B}(s^{'})\;=\frac{12\pi \Gamma^{ee}
\Gamma^{h}}{(s^{'}-M^{2})^{2}
+(\Gamma^{\rm tot} M)^2} ,
\end{equation}
to describe their production,
where $s^{'}$ is the squared actual energy which produces the hadronic events,
$M$ and $\Gamma^{\rm tot}$ are respectively the masses
and total widths of the resonances,
and $\Gamma^{ee}$
and 
$\Gamma^{h}$ are the partial widths
to $e^{+}e^{-}$ channel and to the inclusive hadronic final
states, respectively. 
Assuming that there are no other new structures and effects
except the $\psi(3770)$ resonance and the continuum hadron
production in the energy region
from 3.70 GeV to 3.86 GeV,
we use the pure p-wave Born order Breit-Wigner function
with an energy-dependent total width
to describe the $\psi(3770)$ production
and the $D\bar D$ ($D^0\bar D^0$ and $D^+ D^-$ ) production
from the $\psi(3770)$ decays.
The $\psi(3770)$ resonance shape is taken as
\begin{equation}
\sigma^{\rm B}_{\psi(3770)}(s^{'}) =
    \frac{12 \pi \Gamma^{ee}_{\psi(3770)}
            \Gamma^{\rm tot}_{\psi(3770)}(s^{'})}
{{(s^{'}-M_{\psi(3770)}^2)^2 + 
[M_{\psi(3770)}\Gamma^{\rm tot}_{\psi(3770)}(s^{'})]^2}},
\end{equation}
while the $D^0\bar D^0$, $D^+D^-$ (or $D\bar D$)
resonances shapes are taken as
\begin{equation}
\sigma^{\rm B}_{D\bar D}(s^{'}) =
    \frac{12 \pi \Gamma^{ee}_{\psi(3770)}\Gamma_{D\bar D}(s^{'})}
{{(s^{'}-M_{\psi(3770)}^2)^2 + [M_{\psi(3770)}
   \Gamma^{\rm tot}_{\psi(3770)}(s^{'})]^2}},
\end{equation}
where 
$M_{\psi(3770)}$ and $\Gamma^{ee}_{\psi(3770)}$ are 
the mass and leptonic width of the $\psi(3770)$ resonance, respectively;
$\Gamma_{D\bar D}$ is the partial width 
of $\psi(3770)$ decay into $D \bar D$; 
$\Gamma^{\rm tot}_{\psi(3770)}(s^{'})$ and $\Gamma_{D\bar D}(s^{'})$ are
chosen to be energy dependent,
which are defined as
\begin{equation}
 \Gamma^{\rm tot}_{\psi(3770)}(s^{'})=\Gamma_{D^0\bar D^0}(s^{'})+
          \Gamma_{D^+D^-}(s^{'})+\Gamma_{{\rm non}-D\bar D}(s^{'}),
\end{equation}
where
\begin{eqnarray}
\Gamma_{D^0\bar D^0}(s^{'}) & = &
\Gamma_0~ \theta_{D^0\bar D^0}
           \frac{(p^{}_{D^0})^3} {(p^0_{D^0})^3}
            \frac{1+(rp_{D^0}^0)^2} {1+(rp_{D^0})^2}B_{00}, 
\end{eqnarray}
\begin{eqnarray}
\Gamma_{D^+D^-}(s^{'}) & = &
\Gamma_0~ \theta_{D^+ D^-}
           \frac{(p^{}_{D^+})^3} {(p^0_{D^+})^3}
            \frac{1+(rp_{D^+}^0)^2} {1+(rp_{D^+})^2}B_{+-}, 
\end{eqnarray}
and
\begin{equation}
\Gamma_{{\rm non-}D\bar D}(s^{'}) = \Gamma_0~\left[ 1
   - B_{00}
   - B_{+-}\right],
\end{equation}
where $p^0_D$ and $p_D$ are respectively 
the momenta of the $D$ mesons
produced at the peak of $\psi(3770)$ and at
the actual c.m. energy $\sqrt{s^{'}}$;
$\Gamma_0$ is the total width of
the $\psi(3770)$ at its peak,
$B_{00}=B(\psi(3770)\rightarrow D^0\bar D^0)$ and
$B_{+-}=B(\psi(3770)\rightarrow D^+D^-)$
are the branching fractions for
$\psi(3770)\rightarrow D^0\bar D^0$ and
$\psi(3770)\rightarrow D^+D^-$, respectively,
which are the fitted parameters,
and $r$ is the interaction radius of the $c\bar c$,
which is left free in the fit; $\theta_{D^0\bar D^0}$ and
$\theta_{D^+ D^-}$ are the step
functions to account for the thresholds of the $D^0\bar D^0$ and
$D^+D^-$ production, respectively.

The non-resonant hadronic background shape is taken as
{\small
\begin{eqnarray} 
\nonumber  \sigma^{nrsnt}_h =\int_{0}^{\infty}ds'' G(s,s'')
  \int^1_0 dx  \frac{ R_{uds}(s') {\sigma^B_{\mu^+\mu^-} }(s')} 
  {|1-\Pi(s')|^2} F(x,s) & \\
\nonumber + f_{D\bar D}\left[ (\frac{p_{D^0}}{E_{D^0}})^3\theta_{D^0\bar D^0}
                  + (\frac{p_{D^+}}{E_{D^+}})^3\theta_{D^+ D^-} \right]
 \sigma^B_{\mu^+\mu^-}(s'), 
                               \\
\end{eqnarray}
}
\noindent
\hspace{-0.8mm}with $s'=s(1-x)$, where $x$ is a parameter related to the total 
energy of the emitted photons and $\sqrt{s}$ is the nominal c.m. energy,
$F(x,s)$ is the sampling function~\cite{kuraev},
$1/{|1-\Pi(s(1-x))|}^2$ is the vacuum polarization correction
function~\cite{berends} including the contributions from all 
$1^{--}$ resonances, 
the QED continuum hadron spectrum 
as well as the contributions from the lepton pairs ($e^+e^-$, 
$\mu^+\mu^-$ and $\tau^+\tau^-$)~\cite{zhangdh_gen};
$\sigma^B_{\mu^+\mu^-}(s')$ is the Born cross section 
for $e^+e^-\rightarrow \mu^+\mu^-$, $E_{D^0}$ and $E_{D^+}$ are respectively 
the energies of $D^0$ and $D^+$ mesons
produced at the actual energy $\sqrt{s'}$, 
$f_{D\bar D}$ is a parameter to be fitted,
and $R_{uds}(s')$ is the $R$ value for the light hadron production through
one photon annihilation directly. In the fit we take the $R_{uds}(s')$ as a
constant in the energy region and left it free; 
we fix the $J/\psi$ resonance parameters
at the values given by PDG~\cite{pdg04}.
We also consider the effects of the BEPC energy spread
on the observed cross sections in the fit.
$G(s,s'')$ in Eq. (10) is the Gaussan function
to describe the c.m. energy distribution of the BEPC machine.

The curves in Fig.~\ref{xsct_had_ddbar1} and Fig.~\ref{xsct_had_ddbar2}
show the fits to the data. 
Fig.~\ref{xsct_had_ddbar1}(a) 
shows the observed cross sections with the fit to the data, 
where the points with error show the observed cross sections and
the error is combined from statistical and point-to-point systematic
uncertainties arising from the statistical uncertainties 
in the efficiencies for detection of the
hadronic events and Bhabha events;
the solid line shows the fit to the cross sections and
the dashed line represents the contributions from $J/\psi$, $\psi(2S)$ and
continuum hadron production. 
To examine the contribution from the vacuum
polarization corrections to the Born hadronic cross section due to one
photon annihilation directly, we subtract the contributions of $\psi(2S)$
and $\psi(3770)$ as well as $J/\psi$ from
the observed cross sections to yield the expected cross sections
of the continuum hadron production 
corrected with the radiative effects,
which is given by Eq. (10). 
The squares with error in Fig.~\ref{xsct_had_ddbar1}(b) show the 
yielded-expected cross sections, 
where the errors are the originally absolute errors 
of the totally observed cross sections 
as shown in Fig.~\ref{xsct_had_ddbar1}(a).  
The line 2 in Fig.~\ref{xsct_had_ddbar1}(b) 
shows the fit to the expected cross sections of the continuum hadron
production corrected for the radiative effects.
Fig.~\ref{xsct_had_ddbar2}(a) shows the observed cross sections for the
inclusive hadronic event production, where the contributions from $J/\psi$
and $\psi(2S)$ radiative tails as well as the continuum hadron production
are removed.
Fig.~\ref{xsct_had_ddbar2}(b) and Fig.~\ref{xsct_had_ddbar2}(c)
display the observed cross sections for $D^0\bar D^0$ and $D^+D^-$
production, respectively, where the points with error show the observed
cross sections, while the lines give the fits to the data.
The error is combined from the statistical and point-to-point systematic
arising from the statistical uncertainties in the efficiencies for
detection of the singly tagged $D$ events and Bhabha events.
The fit gives $\chi^2/{\rm n.o.f} = 65.4/64 = 1.02$. 
In the data reduction, the number of the singly tagged 
$D^0$ (or $D^+$) events are removed from the inclusive
hadronic event samples before calculating the hadronic cross section and its
error based on Eq.(1). These make the hadronic event samples and the 
singly tagged $D$ samples be independent.

The results from this fit are
$$B(\psi(3770)\rightarrow D^0\bar D^0) = (46.7 \pm 4.7 \pm 2.3)\%$$
and
$$B(\psi(3770)\rightarrow D^+ D^-) = (36.9 \pm 3.7 \pm 2.8)\%,$$
\noindent
where the first error is statistical and second systematic arising from
uncanceled systematic uncertainties
in the measured $\sigma_{\rm had}$ 
($\sim 2.8\%$)
in $\sigma_{D^0\bar D^0}$ ($\sim 4.1\%$), and
in $\sigma_{D^+ D^-}$ ($\sim 7.0\%$).
Considering the correlation between the two branching fractions obtained
from the fit, we obtain the branching fraction for
$\psi(3770)\rightarrow D\bar D$ to be
$$B(\psi(3770)\rightarrow D\bar D) = (83.6 \pm 7.3 \pm 4.2)\%,$$
which results in the non-$D\bar D$ branching fraction to be
$$B(\psi(3770)\rightarrow non-D\bar D) = (16.4 \pm 7.3 \pm 4.2)\%.$$
The fit gives the BEPC machine energy spread $\sigma_{\rm E_{\rm
BEPC}}=(1.343 \pm 0.029)$ MeV. From the fit we obtain the $R$ value
for the light hadron production due to one photon annihilation
in the energy region between 3.660 GeV and 3.872 GeV to be
          $$R_{uds} = 2.262 \pm 0.054 \pm 0.109,$$
where the first error is statistical and the second systematic arising from
the uncertainty in the measured cross sections of $\sigma_{\rm had}$
($\sim 4.8\%$),
$\sigma_{D^+D^-}$
($\sim 0.4\%$)
and 
$\sigma_{D^0\bar D^0}$
($\sim 0.2\%$).
The fit also gives
the resonance parameters 
of $\psi(3770)$ and $\psi(2S)$ to be 
$M_{\psi(3770)}=3772.2\pm 0.7\pm 0.3$ MeV,
$\Gamma^{\rm tot}_{\psi(3770)}=\Gamma_0=26.9 \pm 2.4 \pm 0.3$ MeV,
$\Gamma^{ee}_{\psi(3770)}=251 \pm 26 \pm 11$ eV;
$M_{\psi(2S)}=3685.5 \pm 0.0 \pm 0.3$ MeV,
$\Gamma^{\rm tot}_{\psi(2S)}=331 \pm 58 \pm 2$ keV,    
$\Gamma^{ee}_{\psi(2S)}=2.330 \pm 0.036 \pm 0.110$ keV,
where the first error is statistical and the second systematic arising from
the uncertainties in the measured $\sigma_{\rm had}$ ($\sim 4.4\%$),
in $\sigma_{D^0\bar D^0}$ ($\sim 4.5\%$), and
in $\sigma_{D^+ D^-}$ ($\sim 7.4\%$).
The measured mass difference between $\psi(3770)$ and $\psi(2S)$
is $\Delta M=86.6 \pm 0.7$ MeV. 
The measured values of the $\psi(3770)$ resonance parameters
yield the cross section for $\psi(3770)$ production at its peak to
be $\sigma^{\rm prd}_{\psi(3770)}=(9.63 \pm 0.66 \pm 0.35)$ nb,
corresponding the observed cross section 
$\sigma^{\rm obs}_{\psi(3770)}=(6.94 \pm 0.48 \pm 0.28)$ nb,
which is consistent within error with 
$\sigma^{\rm obs}_{\psi(3770)}=(8.12 \pm 1.56)$ nb at $\psi(3770)$ peak
obtained based on the PDG
$\psi(3770)$ resonance parameters. 

The measured branching fractions yield the ratio of the partial widths
$\Gamma_{D^0\bar D^0}/\Gamma_{D^+D^-} = 1.27\pm 0.12 \pm 0.08$, which 
agrees with the prediction of 1.36 by Eichten et al.~\cite{eichten}
and is in good agreement with $1.41 \pm 0.23\pm 0.11$ measured by
BES~\cite{bes2_xsctddbar_abslt}.
The measured widths of the resonances yield 
the leptonic branching fractions to be 
$BF(\psi(3770)\rightarrow e^+e^-)=(0.93\pm 0.06 \pm 0.03) \times 10^{-5}$ and
$BF(\psi(2S)\rightarrow e^+e^-)=(0.704 \pm 0.122 \pm 0.033)\%$. 

We find that the continuum background shape affect the measured 
total and leptonic widths of the resonances
from the line-shape analysis.
If we take the non-resonant hadronic background shape as
{\small
\begin{eqnarray}
\sigma^{nrsnt}_h = & 
\nonumber h~\sigma^B_{\mu^+\mu^-}(s') \hspace{50mm} & \\
\hspace{0mm}
\nonumber + & f_{D\bar D}\left[(\frac{p_{D^0}}{E_{D^0}})^3\theta_{D^0\bar D^0}
                 +(\frac{p_{D^+}}{E_{D^+}})^3\theta_{D^+ D^-} \right]
 \sigma^B_{\mu^+\mu^-}(s'), \\
\end{eqnarray}
}
\noindent
\hspace{-1.2mm}in fitting the data (where $h$ is 
a parameter in the fit), we would obtain 
$\Gamma^{\rm tot}_{\psi(2S)}=290 \pm 59 \pm 5$ keV,
$\Gamma^{ee}_{\psi(2S)}=2.378 \pm 0.036 \pm 0.103$ keV, 
$\Gamma^{\rm tot}_{\psi(3770)}=27.3 \pm 2.5 \pm 1.1$ MeV and
$\Gamma^{ee}_{\psi(3770)}=256 \pm 27 \pm 13$ eV,
and almost unchanged the measured masses of the resonances.
This fit yields $\chi^2/{\rm n.o.f} = 75.3/64 = 1.18$.
These indicate that the vacuum polarization corrections 
to the Born order cross sections for the continuum hadron production
can not be ignored
in 
precisely
measuring the resonance parameters of the narrow
resonances like $J/\psi$ and $\psi(2S)$ as well as $\Upsilon(1S)$ 
etc. in $e^+e^-$ cross section scan experiments. 
Ignoring the effects of the vacuum polarization corrections on the continuum hadron
production cross sections in analysis of the cross section scan data
taken in the $\psi(2S)$ resonance region would decrease the $\psi(2S)$ total
width by about 40 keV.

In summary, 
we measured the branching fractions 
for $\psi(3770)\rightarrow D^0\bar D^0,D^+D^-$ and
$\psi(3770)\rightarrow {\rm non}-D\bar D$ for the first time 
and measured the resonance parameters of $\psi(3770)$ and
$\psi(2S)$ with improved precision on $\psi(3770)$ resonance parameter
and with a comparable precision to the current PDG world averages
on $\Gamma^{ee}_{\psi(2S)}$. With the same data samples, we
also measured the $R$ value for the continuum light hadron production 
with a precision of about $\pm 5\%$.
From this analysis we directly observed the effects of the vacuum polarization
on the observed cross sections of the continuum hadron production in the
neighborhood of the $\psi(2S)$ resonance. 
\vspace{6mm}

   The BES collaboration thanks the staff of BEPC for their hard efforts.
One of the authors (G. Rong) would like to thank Professor F. Porter 
for useful discussions about the
measurement of $B(\psi(3770)\rightarrow D^0\bar D^0,~D^+D^-,~D \bar D)$.
This work is supported in part by the National Natural Science Foundation
of China under contracts
Nos. 19991480,10225524,10225525, the Chinese Academy
of Sciences under contract No. KJ 95T-03, the 100 Talents Program of CAS
under Contract Nos. U-11, U-24, U-25, and the Knowledge Innovation Project
of CAS under Contract Nos. U-602, U-34(IHEP); by the
National Natural Science
Foundation of China under Contract No.10175060(USTC),and
No.10225522(Tsinghua University).

\end{document}